\documentclass[superscriptaddress,reprint,amssymb, nobibnotes, aps, pra]{revtex4-2}

\usepackage{graphicx}
\usepackage{amsmath}
\usepackage{float}
\usepackage{braket}
\usepackage{bibentry}

\usepackage{color,soul}
\usepackage{appendix}

\begin{document}

\title{Contactless Continuous-Variable Quantum Optical State Conditioning With Classical MmWave Phase}

\author{Niloy Ghosh}
\affiliation{Netherlands Institute for Radio Astronomy (ASTRON), 7991 PD Dwingeloo, The Netherlands}

\author{Sarang Pendharker}
\affiliation{Department of Electronics $\&$ Electrical Communication Engineering, Indian Institute of Technology Kharagpur, W. Bengal, 721302, India.}

\begin{abstract}
This paper establishes the Heisenberg's picture framework for seamless mmWave-to-photonic data transduction. Based on this framework, contactless modulation of continuous-variable (CV) quantum optical states with digitally modulated classical mmWave beams is shown for the first time. Our analysis reveals the direct influence of mmWave phase variation on the Wigner-space evolution of modulated optical coherent and squeezed states. Furthermore, the analysis indicates that classical mmWave phase not only governs the displacement but also modulates the quadrature uncertainty of squeezed states, revealing a novel technique for embedding information onto multiple quantum-optical degrees of freedom. This work builds the theoretical and analytical foundation for wireless conditioning of quantum light without conventional on-chip metallic interconnects and complex electronic control circuitry, thereby opening a key pathway toward seamless bridging of integrated quantum photonics with emerging mmWave or THz communication technology.
\end{abstract}

\maketitle

\section{Introduction}

Quantum information processing promises to outperform the classical counterpart in applications ranging from secure communication to solving complex optimization problems \cite{kim2023evidence, herman2023quantum, daley2022practical}. Several matter-based platforms like superconducting transmons, trapped ions, and semiconductor quantum dots, are being actively explored to practically enable quantum information processing \cite{cao2023generation, bardin2021microwaves, krutyanskiy2023entanglement, mcjunkin2022sige}. Despite significant progress, these platforms remain highly susceptible to thermal noise and require cryogenic operation. This increases system complexity, energy consumption, cost, and limits scalability \cite{zhai2024development}. In contrast, photonic platforms are relatively immune to thermal noise at room temperature, and additionally offer inherent mobility \cite{furusawa2011quantum}. Together with rapid advances in integrated photonics, these advantages have made photonics one of the most versatile technological platforms for implementing quantum computing and networking \cite{bogaerts2020programmable}.

\begin{figure}[h!]
    \centering 
    \includegraphics[width=\linewidth]{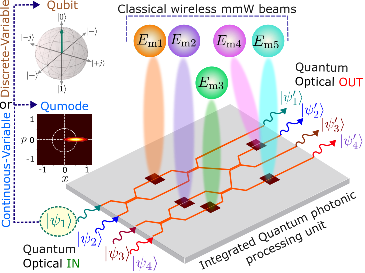}
    \caption{Possible system architecture for contactless addressing of multiple quantum optical states through classical mmWave or sub-THz beams.}
    \label{Figure1_0}
\end{figure}

Achieving quantum supremacy partly hinges on the capability to deterministically address multiple quantum optical states on the same platform. Existing integrated photonic platforms typically rely upon classical RF signals distributed through on-chip metallic control lines \cite{qiang2018large, larsen2025integrated}. However, these interconnects suffer from crosstalk, electromagnetic interference, and ohmic losses, that degrade the overall quantum control fidelity \cite{pozar2021microwave}. As the number of quantum channels increase, the RF distribution network becomes more dense, posing a major bottleneck in terms of footprint, power consumption, and scalability. A promising alternative is contactless  control, where on-chip metallic interconnects are replaced by highly directive wireless mmWave or THz beams, as illustrated in Fig.~\ref{Figure1_0}. By exploiting existing spatial and frequency-division multiplexing techniques, multiple qubits \cite{asavanant2022optical} or qumodes \cite{hajomer2024long} can be simultaneously addressed without dense electrical routing, thereby enabling a more scalable system architecture. 

Realizing contactless quantum state addressing requires a rigorous theoretical framework for modeling the interaction between quantum optical states and classical wireless mmWave or THz signals. Existing models are largely restricted to continuous-wave analog microwave modulation \cite{capmany2010quantum, horoshko2018quantum, miroshnichenko2017algebraic}. We previously investigated contactless modulation of classical optical waves using digitally modulated mmWave and sub-THz beams \cite{Ghosh_2021, 10107736, 10535154, ghosh2023impact}, and later extended this to  quantum optical state encoding within the Schrodinger's picture \cite{10.1063/5.0226097}. However, a general framework for contactless conditioning of continuous-variable quantum optical states is still missing. Addressing this gap requires a Heisenberg-picture formulation of seamless mmWave-to-optical data transduction, which constitutes the primary goal of this work.


This paper models the quantum mechanical framework for seamless classical-mmWave to quantum-optical data transduction in Heisenberg's picture. The developed model is then extended to simulate contactless conditioning of CV quantum photonic states. A direct link between the baseband variation of the classical mmWave signal and the Wigner-space evolution of the correspondingly modulated quantum optical state is established, providing a unified analytical description for wireless control of quantum light. Furthermore, mapping of classical mmWave 4-PSK and 8-PSK constellations onto the Wigner distributions of weak optical coherent and squeezed states is shown. The analysis also reveals that the classical mmWave phase engineers the quadrature uncertainty of the modulated squeezed states, allowing simultaneous distribution of information onto multiple quantum optical parameters. The reported contributions provide a foundation for replacing conventional on-chip metallic RF interconnects and associated electronic control circuitry in integrated quantum photonic platforms with highly directive wireless mmWave or THz interfaces, thereby enhancing the scalability of such systems. 

The remainder of the paper is organized as follows. Section~II develops the quantum mechanical framework for contactless modulation of quantum optical quadrature operator with digitally encoded classical mmWave beams. Section~III and IV demonstrate contactless conditioning of Wigner-space distribution of weak optical states and squeezed states with classical mmWave constellations, respectively. Additionally, Section~IV demonstrates quadrature fluctuation and uncertainty modulation of optical squeezed-states with classical mmWave phase.

\section{Quantum mechanical framework for photonic quadrature operator encoding}

\subsection{Quantum photonic Hamiltonian modulation}

\begin{figure}[h!]
    \centering 
    \includegraphics[width=\linewidth]{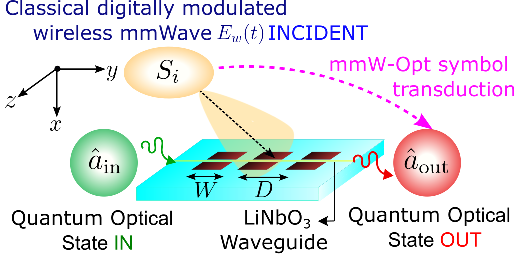}
    \caption{Seamless classical-mmWave to quantum-optical symbol transduction in Heisenberg's picture.}
    \label{Figure1}
\end{figure}

Figure~\ref{Figure1} illustrates the schematic of a seamless mmWave-to-optical (M-O) converter. The converter is essentially an antenna-integrated electro-optic modulator comprising of a Lithium Niobate (LiNbO$_3$) optical waveguide channelized through a centrally slotted metallic patch antenna array. The waveguide present in the converter supports a $z$-polarized 
electromagnetic mode associated with a quantum photonic state along propagating in the $y$-direction. In the absence of mmWave reception, the free Hamiltonian $\hat{H}_{\mathrm{op}}$ of the photonic state resembles that of a quantum harmonic oscillator \cite{bransden2000quantum}. The free Hamiltonian $\hat{H}_{\mathrm{op}}$ of the quantum photonic state can be expressed as follows \cite{khrennikov2012quantization},
\begin{equation}
 \hat{H}_{\mathrm{op}}=\ \hbar \omega_{\mathrm{op}} \bigg( \hat{a}^{\dagger}\hat{a} + \frac{1}{2} \bigg)
\label{Eq0}
\end{equation}
where $\hbar$ is the reduced Planck's constant, and $\omega_{\mathrm{op}}$ is the frequency of the electromagnetic-field associated with the quantum photonic state. Also, $\hat{a}$ and $\hat{a}^{\dagger}$ are the associated field amplitude operator and it adjoint, respectively \cite{fox2006quantum}.

We then consider a $z$-polarized classical mmWave signal $E_{\mathrm{m}}(t)$ of the following form to be incident on the converter,
\begin{equation}
E_{\mathrm{m}}(t) = a_i |E_\mathrm{m}| \sin{(\omega_{\mathrm{m}}t + b_i)}
\label{Eq1}
\end{equation}
where $|E_\mathrm{m}|$ and $\omega_{\mathrm{m}}$ are the E-field strength and frequency of $E_{\mathrm{m}}(t)$, respectively. Here, the received mmWave signal is considered to be digitally modulated, encapsulating the digital baseband symbol $S_i$. In general, $S_i$ can be any possible combination of baseband symbol amplitude $a_i \in [-1,1]$ and phase $b_i \in [-180^{\circ}, 180^{\circ}]$ \cite{haykin1988digital}. 

Upon receiving $E_{\mathrm{m}}(t)$, a strong $z$-polarized mmWave E-field $E_{\mathrm{s}}(t)$ is subsequently induced in the slotted regions of the individual patch antennas present in the converter \cite{wijayanto2013electrooptic}. The slot-induced mmWave E-field $E_{\mathrm{s}}(t)$ is essentially the enhanced version of $E_{\mathrm{m}}(t)$, as indicated by the following equation \cite{park2015free},
\begin{equation}
E_{\mathrm{s}}(t)=e E_{\mathrm{m}}(t)
\label{Eq2}
\end{equation}
where the parameter $e$ represents the slot-field enhancement factor \cite{Ghosh_2021}. Due to the induction of $E_{\mathrm{s}}(t)$ in the slotted regions of the antennas, the refractive index $n_{\mathrm{op}}$ of the under-routed waveguide segments get perturbed. This is governed by Pockels electro-optic effect \cite{zhu2021integrated}. The perturbed optical refractive index $n_{\mathrm{op}}(t)$ is an instantaneous function of the received classical mmWave signal, as indicated below,
\begin{equation}
n_{\mathrm{op}}(t) = n_{\mathrm{op}}-\frac{n_{\mathrm{op}}^{2} r_{33} e}{2} E_{\mathrm{m}}(t)
\label{Eq3}
\end{equation}
where $r_{33}$ is the Pockels electro-optic coefficient of LiNbO$_3$. Consequently, the Hamiltonian $\hat{H}_{\mathrm{op}} (t)$ of the quantum photonic state gets modulated by the received classical mmWave signal, as indicated below,
\begin{equation}
 \hat{H}_{\mathrm{op}} (t)=\ \hbar \omega_{\mathrm{op}} \bigg( \hat{a}^{\dagger}\hat{a} + \frac{1}{2} \bigg) \bigg(1 - \frac{n_{\mathrm{op}}^{2} r_{33} e }{2}  E_{\mathrm{m}}(t) \bigg) 
\label{Eq4}
\end{equation}


The upcoming subsection investigates the impact of classical mmWave-governed Hamiltonian modulation on the quantum photonic quadrature operators.


\subsection{Classical mmWave-governed quantum photonic quadrature operator encoding}

The phase-space distribution of a quantum state is governed by the quadrature operator basis $(\hat{x}, \hat{p})$. The quadrature operators are in turn governed by the field amplitude operators $(\hat{a}, \hat{a}^{\dagger})$. Therefore, to determine the effect of Hamiltonian modulation on the phase-space distribution of a quantum state, the corresponding evolution of the field amplitude operators must first be derived. The evolution of the field amplitude operator $\hat{a}$ associated with the modulated quantum photonic state can be derived from the below-mentioned Heisenberg's equation \cite{gerry2023introductory},
\begin{equation}
\frac{d \hat{a}(t)}{dt} = \frac{j}{\hbar} [\hat{H}_{\mathrm{op}}(t), \hat{a}(t)] 
\label{Eq5}
\end{equation}
Substituting the expression of the modulated Hamiltonian $\hat{H}_{\mathrm{op}}(t)$ from Eq.~(\ref{Eq4}) into the above equation, and solving the resultant commutation, we arrive at the following differential equation,
\begin{equation}
\frac{d \hat{a}(t)}{dt}  = -j \omega_{\mathrm{op}} \bigg(1-  \frac{n^{2}_{\mathrm{op}} r_{33} e  }{2} E_{\mathrm{m}}(t)\bigg) \hat{a}(t)
\label{Eq6}
\end{equation}
It must be reiterated that the quantum photonic state is modulated only while propagating through the waveguide segments routed under the individual antennas in the array. The modulation time-window for the waveguide segment under the $n^{\mathrm{th}}$-antenna spans from $T^{n}_{\mathrm{in}}$=$(t + (N-1) T_D)$ to $T^{n}_{\mathrm{out}}$=$(t + (N-1) T_D + T_W)$. It must be noted that $T^{n}_{\mathrm{in}}$ and $T^{n}_{\mathrm{out}}$ represent the time instants at which the photonic state enters and exits the waveguide segment under the $n^{\mathrm{th}}$ antenna, respectively. Here, $T_{W}$=$(n_{\mathrm{op}} W/c)$ and $T_D$=$(n_{\mathrm{op}} D/c)$ is the time taken by the photonic state to travel distance $W$ (antenna width) and $D$ (array periodicity), respectively, in the optical waveguide. So, the net evolution of $\hat{a}(t)$ can be derived by integrating Eq.~(\ref{Eq6}) within the time-window spanning from $T^{1}_{\mathrm{in}}$=$t$ to $T^{N}_{\mathrm{out}}$=$(t + (N-1) T_D + T_W)$, leading upto the following summation-over-integral,
\begin{equation}
\int_{T^{1}_{\mathrm{in}}}^{T^{N}_{\mathrm{out}}} \frac{d \hat{a}(t')}{\hat{a}(t')} = -j\omega_{\mathrm{op}} \sum_{n=1}^{n=N} \int_{T^{n}_{\mathrm{in}}}^{T^{n}_{\mathrm{out}}} \bigg(1- \frac{n^{2}_{\mathrm{op}} r_{33} e}{2} E_\mathrm{m}(t')   \bigg)\ dt' 
\label{Eqq7}
\end{equation}
Solving the above equation, we get,
\begin{equation}
   \hat{a}_\mathrm{out}(t)  = \hat{a}_\mathrm{in}(t) e^{-j \theta_i(t)} 
\label{Eq8}
\end{equation}
where $\hat{a}_{\mathrm{in}}(t)$ and $\hat{a}_{\mathrm{out}}(t)$ are the time-dependent field amplitude operators of the unmodulated photonic state at the array input and that of the modulated photonic state at the array output, respectively. It must be noted that we have ignored the trivial optical propagation phase term $e^{-jk_{\mathrm{op}}(W+(N-1)D)}$ that would otherwise appear in Eq.~(\ref{Eq8}), with $k_{\mathrm{op}}$ being the free-space optical phase-constant. 

It is interesting to note from Eq.~(\ref{Eq8}) that field amplitude operator
$\hat{a}_\mathrm{out}(t)$ of the modulated photonic state gets encoded with the phase $\theta_i(t)$ as a consequence of Hamiltonian modulation. The modulated optical phase $\theta_i(t)$ is directly related to the received classical mmWave signal, as follows,
\begin{equation}
    \theta_i (t) = a_i \delta \theta |E_\mathrm{m}| \cos{(\omega_{\mathrm{m}}t+b_i+r)}
\label{Eq9}
\end{equation}
where $\delta \theta$ refers to the electro-optic modulation-depth and $r$ refers to the effective wireless constellation rotation angle, that can be expressed as follows \cite{10107736},
\begin{subequations}
    \begin{equation}
        \delta \theta = \delta \theta_{\mathrm{o}}  \sin{\bigg(\frac{k_{\mathrm{m}}  n_{\mathrm{op}} W}{2} \bigg)}\ \frac{\sin{(N k_{\mathrm{m}}  n_{\mathrm{op}} D/2)}}{\sin{(k_{\mathrm{m}}  n_{\mathrm{op}} D/2)}}
    \end{equation}
    \begin{equation}
       r= \frac{k_{\mathrm{m}}  n_{\mathrm{op}} W}{2} - \frac{\pi}{2}
    \end{equation}
\label{Eq10}
\end{subequations}
where $k_{\mathrm{m}}$ is the free-space mmWave phase-constant, and $\delta \theta_{\mathrm{o}}$ is the peak modulation-depth achievable per antenna which can be expressed as,
\begin{equation}
  \delta \theta_{\mathrm{o}}= -\frac{\omega_{\mathrm{op}} n^{2}_{\mathrm{op}} r_{33} e}{\omega_{\mathrm{m}}} 
\label{Eq11}
\end{equation}

By substituting the expression of $\theta_i(t)$ from Eq.~(\ref{Eq9}) into Eq.~(\ref{Eq8}), the resultant form of  $\hat{a}_{\mathrm{out}}(t)$ can be expanded using Jacobi Anger identity as follows,
\begin{equation}
  \hat{a}_{\mathrm{out}}(t) =  \sum_{s=-\infty}^{\infty}  J_s(\rho)\ \hat{a}_{\mathrm{in}}(t)\ e^{-js \phi_i}  e^{-js\omega_{\mathrm{m}}t} 
\label{Eq15_0}
\end{equation}
where $J_s(\rho)$ is the Bessels function of $s^{\mathrm{th}}$ kind. The corresponding argument term $\rho$ is related to baseband symbol amplitude $a_i$, modulation-depth $\delta \theta$, and mmWave E-field strength $|E_{\mathrm{m}}|$ as follows,
\begin{equation}
  \rho = a_i\delta \theta |E_{\mathrm{m}}|
\label{Eq15_00}
\end{equation}
Also, the phase term $\phi_i$ in Eq.~(\ref{Eq15_0}) is related to the baseband symbol phase $b_i$ and constellation-rotation angle $r$ as follows,
\begin{equation}
  \phi_i = b_i+r-\frac{\pi}{2}
\label{Eq15_01}
\end{equation}

In Heisenberg's picture,  $\hat{a}_{\mathrm{in}}(t)$ can be represented as $\hat{a}_{\mathrm{in}}^{0} e^{-j\omega_{\mathrm{op}}t}$. Here, $\hat{a}_{\mathrm{in}}^{0}$ is the time-independent field amplitude operator of the unmodulated quantum photonic state at the array input that naturally evolves at optical carrier frequency $\omega_{\mathrm{op}}$, represented by the time-harmonic term $e^{-j\omega_{\mathrm{op}}t}$. By substituting $\hat{a}_{\mathrm{in}}(t)$=$\hat{a}_{\mathrm{in}}^{0} e^{-j\omega_{\mathrm{op}}t}$ in Eq.~(\ref{Eq15_0}), the resultant expression of $\hat{a}_{\mathrm{out}}(t)$ comes out to be,
\begin{equation}
  \hat{a}_{\mathrm{out}}(t) =  \sum_{s=-\infty}^{\infty}  J_s(\rho)\ \hat{a}_{\mathrm{in}}^{0}\ e^{-js \phi_i}  e^{j(\omega_{\mathrm{op}}+s\omega_{\mathrm{m}})t} 
\label{Eq15_001}
\end{equation}
The above equation indicates that the field amplitude operator $\hat{a}_{\mathrm{in}}^{0}$ of the unmodulated photonic state splits into different frequency sidebands upon being modulated by the classical mmWave signal. Therefore, the time-dependent field amplitude operator $\hat{a}_{\mathrm{out}}(t)$ of the modulated state can be expressed as the summation of all sideband components as follows,
\begin{equation}
  \hat{a}_{\mathrm{out}}(t) =  \sum_{s=-\infty}^{\infty}  \hat{a}^{s}_{\mathrm{out}} e^{-j(\omega_{\mathrm{op}}+s\omega_{\mathrm{m}})t} 
\label{Eq15_1}
\end{equation}
where $\hat{a}^{s}_{\mathrm{out}}$ is the time-independent field-amplitude component associated with the $s^{th}$-sideband of the modulated quantum state. This is related to the field amplitude operator $\hat{a}^{0}_{\mathrm{in}}$ of the unmodulated quantum state at the converter input through the following relationship,
\begin{equation}
  \hat{a}^{s}_{\mathrm{out}} = J_s(\rho)\  \hat{a}^{0}_{\mathrm{in}}\ e^{-js \phi_i}
\label{Eq15_2}
\end{equation}
It must be reiterated that the sidebands arise due to the Hamiltonian modulation of the quantum state in the converter. Even though multiple sidebands may be present in general, we are primarily interested in the first-order sideband of the modulated quantum state. The time-independent field amplitude operator $\hat{a}^{+1}_{\mathrm{out}} $ associated with the first-order sideband ($s$=1) evolving at frequency $(\omega_{\mathrm{op}}+\omega_{\mathrm{s}})$ can be expressed as,
\begin{equation}
  \hat{a}^{+1}_{\mathrm{out}} = J_1(\rho)\  \hat{a}_{\mathrm{in}}^{0}\ e^{-j \phi_i}
\label{Eq15}
\end{equation}
Likewise, the adjoint operator $\hat{a}^{+1}_{\mathrm{out}}$ can be expressed as,
\begin{equation}
  (\hat{a}^{+1}_{\mathrm{out}})^{\dagger} = J_1(\rho)\  \hat{a}^{0\dagger}_{\mathrm{in}}\ e^{j \phi_i}
\label{Eq15_part2}
\end{equation}
Our target now is to derive the time-independent quadrature operators associated with the first-order sidebands of modulated quantum photonic state. The quadrature operators $[\hat{x}^{+1}_{\mathrm{out}}, \hat{p}^{+1}_{\mathrm{out}}]$ are related to the field-amplitude operators $[\hat{a}^{+1}_{\mathrm{out}},  (\hat{a}^{+1}_{\mathrm{out}})^{\dagger}]$
through the following relation \cite{fox2006quantum},
\begin{subequations}
    \begin{equation}
       \hat{x}^{+1}_{\mathrm{out}} = \frac{\hat{a}^{+1}_{\mathrm{out}} +(\hat{a}^{+1}_{\mathrm{out}})^{\dagger}}{2}
    \end{equation}
    \begin{equation}
      \hat{p}^{+1}_{\mathrm{out}} = \frac{\hat{a}^{+1}_{\mathrm{out}} -(\hat{a}^{+1}_{\mathrm{out}})^{\dagger}}{2j}
    \end{equation}
\label{Eq15_part3}
\end{subequations}

Substituting the derived expressions of $\hat{a}^{+1}_{\mathrm{out}}$ and $ (\hat{a}^{+1}_{\mathrm{out}})^{\dagger}$ from Eq.~(\ref{Eq15}) and Eq.~(\ref{Eq15_part2}), respectively, into Eq.~(\ref{Eq15_part3})(a) and Eq.~(\ref{Eq15_part3})(b), the quadrature operators $[\hat{x}^{+1}_{\mathrm{out}}, \hat{p}^{+1}_{\mathrm{out}}]$ of the modulated state at the converter output can be related to  quadrature operators $[\hat{x}^{0}_{\mathrm{in}}, \hat{p}^{0}_{\mathrm{in}}]$ of the unmodulated state at the converter input as follows,
\begin{subequations}
\begin{equation}
  \hat{x}^{+1}_{\mathrm{out}} = J_1(\rho) \cos{(\phi_i)}\ \hat{x}^{0}_{\mathrm{in}} + J_1(\rho) \sin{(\phi_i)}\ \hat{p}^{0}_{\mathrm{in}}
\end{equation}
\begin{equation}
  \hat{p}^{+1}_{\mathrm{out}} = -J_1(\rho) \sin{(\phi_i)}\ \hat{x}^{0}_{\mathrm{in}} + J_1(\rho) \cos{(\phi_i)}\ \hat{p}^{0}_{\mathrm{in}}
\end{equation}
 \label{Eq17}
\end{subequations}
where the quadrature operators $[\hat{x}^{0}_{\mathrm{in}}, \hat{p}^{0}_{\mathrm{in}}]$ are associated with the field-amplitude operators $[\hat{a}^{+0}_{\mathrm{in}},  (\hat{a}^{+0}_{\mathrm{in}})^{\dagger}]$ of the unmodulated quantum photonic state as follows,
\begin{subequations}
\begin{equation}
       \hat{x}^{0}_{\mathrm{out}} = \frac{\hat{a}^{0}_{\mathrm{out}} +(\hat{a}^{0}_{\mathrm{out}})^{\dagger}}{2}
    \end{equation}
    \begin{equation}
      \hat{p}^{0}_{\mathrm{out}} = \frac{\hat{a}^{0}_{\mathrm{out}} -(\hat{a}^{0}_{\mathrm{out}})^{\dagger}}{2j}
    \end{equation}
 \label{Eq17_2}
\end{subequations}

From Eq.~(\ref{Eq17}), it can be concluded that the quadrature operators of the modulated quantum photonic state are governed by the modulation parameters $\rho$ and $\phi_i$. It can be reiterated from Eq.~(\ref{Eq15_00}) that $\rho$ is a function of the modulation-depth $\delta \theta$ and baseband symbol amplitude $a_i$. Additionally, it can be reiterated from Eq.~(\ref{Eq15_01}) that $\phi_i$ is a function of baseband symbol phase $b_i$. It can therefore be concluded that the seamless converter mediates mmWave-photonic symbol transduction, due to which the quadrature-operators of the modulated quantum photonic state gets encoded with the baseband symbol $S_i$=[$a_i, b_i$] encapsulated within the modulating classical mmWave signal. 

The efficiency of mmWave-photonic symbol transduction is governed by the modulation-depth. This can be achieved by optimally choosing the individual width of the antennas as $W$=$(\pi/2k_{\mathrm{m}} n_{\mathrm{op}})$ and the array periodicity as $D$=$(\pi/k_{\mathrm{m}} n_{\mathrm{op}})$ \cite{10.1063/5.0226097}. In an optimally designed N-element array, the modulation-depth linearly upscales with the number of cascaded antennas as $\delta \theta$=$|N \delta \theta_{\mathrm{o}}|$. Also, the constellation-rotation angle under optimal design conditions is $r$=$0^{\circ}$. In this paper, we have chosen the optimum antenna width $W$=2.9mm and array-periodicity as $D$=5.8mm for 30GHz mmWave operating frequency.

In the upcoming sections, we will investigate how the phase-space distribution of different quantum photonic states like coherent and squeezed states can be conditioned through seamless mmWave-to-photonic symbol transduction.

\section{Conditioning optical coherent states}

A coherent state $\ket{\alpha}$ can be defined as a displaced vacuum state, and mathematically described as follows \cite{fox2006quantum},
\begin{equation}
  \ket{\alpha} = \hat{D}(\alpha) \ket{0}
\label{Eq20_0}
\end{equation}
where $\hat{D}(\alpha)$=$\exp{(\hat{a}^{\dagger} \alpha - \hat{a} \alpha^{*})}$ is the displacement operator, and $\ket{0}$ represents a vacuum state. Also,  $\alpha$ is the mean complex amplitude of the electric-field associated with the optical coherent state $\ket{\alpha}$. The mean complex amplitude $\alpha$ can therefore be expressed as $|\alpha| e^{-j\phi}$, where $|\alpha|$ is the mean magnitude and $\phi$ is the mean phase of the E-field associated with $\ket{\alpha}$.

We next consider a weak optical coherent state $\ket{\alpha}$ to be modulated by the received classical mmWave signal. The phase-space distribution of the first-order sideband of the modulated optical coherent state $\ket{\alpha}$ can be described by the following Wigner function,
\begin{equation}
  W(x,p) = \frac{2}{\pi} \exp{\big(-2((x-\braket{\hat{x}^{+1}_{\mathrm{out}}}_{})^{2}+(p-\braket{\hat{p}^{+1}_{\mathrm{out}}}_{})^{2}})\big)
\label{Eq19}
\end{equation}
where $\braket{\hat{x}^{{+1}_{\mathrm{out}}}}$ and $\braket{\hat{p}^{+1}_{\mathrm{out}}}$ is the mean $\hat{x}$ and $\hat{p}$ associated with the first-order sideband of the modulated coherent state $\ket{\alpha}$, respectively. The mean of the quadrature operator ${\hat{x}^{+1}_{\mathrm{out}}}$ can be derived from Heisenberg's operator transformation picture as follows,
\begin{equation}
\begin{split}
\braket{\hat{x}^{+1}_{\mathrm{out}}} & =  \bra{\alpha} \hat{x}^{+1}_{\mathrm{out}} {\ket{\alpha}} \\
  & = \Re{\{\alpha\}} J_1(\rho)\cos{\phi_i} + \Im{\{\alpha\}} J_1(\rho)\sin{\phi_i}
\end{split}
\label{Eq21}
\end{equation}
Similarly, the mean of the quadrature operator $\hat{p}^{+1}_{\mathrm{out}}$ can be derived to be,
\begin{equation}
\begin{split}
\braket{\hat{p}^{+1}_{\mathrm{out}}} & =  \bra{\alpha} \hat{p}^{+1}_{\mathrm{out}} {\ket{\alpha}} \\
  & =-\Re{\{\alpha\}} J_1(\rho)\sin{\phi_i} + \Im{\{\alpha\}} J_1(\rho)\cos{\phi_i}
\end{split}
\label{Eq20}
\end{equation}

\begin{figure}
    \centering
    \includegraphics[width=\linewidth]{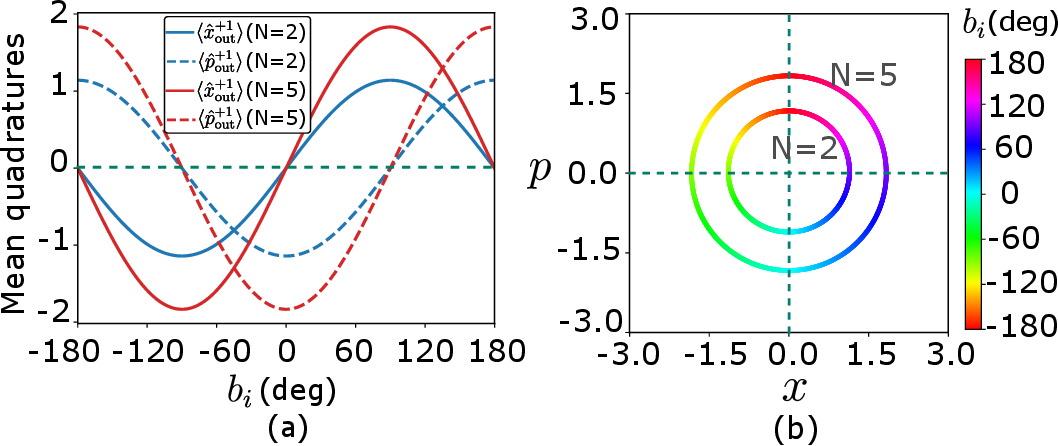}
    \caption{(a) Variation in mean quadratures $\braket{\hat{x}^{+1}_{\mathrm{out}}}$ and $\braket{\hat{p}^{+1}_{\mathrm{out}}}$, and (b) Mean locus of the modulated optical coherent-state composed of a mean photon number of $\tilde{n}_{ph}$=4 in the phase-space observed by sweeping the baseband symbol phase from $b_i$=$-180^{\circ}$ to $b_i$=$180^{\circ}$, when the number of optimally cascaded antennas in the converter are N=2 and N=5.}
    \label{Figure2}
\end{figure}

Substituting the derived expressions of quadrature operators above into Eq.~(\ref{Eq19}), the Wigner function $W(x,p)$ of the modulated coherent-state can be expressed in terms of the modulation parameters $\rho$ and $\phi_i$ as follows,
\begin{widetext}
\begin{equation}
 W(x,p) =  \frac{2}{\pi} \exp{\big(-2((x-\Re{\{\alpha\}} J_1(\rho)\cos{\phi_i} - \Im{\{\alpha\}} J_1(\rho)\sin{\phi_i})^{2}+(p+\Re{\{\alpha\}} J_1(\rho)\sin{\phi_i} - \Im{\{\alpha\}} J_1(\rho)\cos{\phi_i})^{2})\big)}
\label{Eq25}
\end{equation}
\end{widetext}
Since $\rho$ and $\phi_i$ are governed by the baseband symbol, Eq.~(\ref{Eq25}) indicates that Wigner function of the modulated coherent state is also governed by it. This implies that the baseband symbol $S_i$=$(a_i, b_i)$ encoded in the modulating classical mmWave signal influences the phase-space distribution of the modulated quantum photonic state. Therefore, tuning the baseband amplitude, and more interestingly the phase can lead to conditioning of a quantum optical state. Figure~\ref{Figure2} further substantiates this argument. Figure~\ref{Figure2}(a) shows the variation in mean quadratures $\braket{\hat{x}^{+1}_{\mathrm{out}}}$ and $\braket{\hat{p}^{+1}_{\mathrm{out}}}$ of the modulated optical coherent-state observed by sweeping the baseband symbol phase from $b_i$=$-180^{\circ}$ to $b_i$=$ 180^{\circ}$. It was previously shown in Eq.~(\ref{Eq15_01}) that the modulated optical phase $\phi_i$ and the baseband symbol phase $b_i$ share a linear relationship.  Therefore, it can be inferred from Eq.~(\ref{Eq21}) and Eq.~(\ref{Eq20}) that $\braket{\hat{x}^{+1}_{\mathrm{out}}}$ and $\braket{\hat{p}^{+1}_{\mathrm{out}}}$ are sinusoidal functions of $b_i$, as shown in Fig.~\ref{Figure2}(a). On the other hand, Fig.~\ref{Figure2}(b) traces the mean locus of the modulated coherent-state in the phase-space as $b_i$ is swept from $b_i$=$-180^{\circ}$ to $180^{\circ}$. It can be noted from Fig.~\ref{Figure2}(b) that shifting the baseband phase $b_i$ embedded in the modulating classical mmWave signal translates to a rotation operation of the correspondingly modulated optical coherent state in the phase-space. This can be attributed to the mean quadratures of the modulated coherent-state being sinusoidal functions of $b_i$ shown in Fig.~\ref{Figure2}(a).

Furthermore, it can also be noted that the mean radial distance of the modulated coherent-state from the origin of the phase-space increases as the number of optimally cascaded antennas in the converter is increased from N=2 to N=5. This is expected because the mean radial distance from the origin of phase-space $d$=$|\sqrt{\braket{\hat{x}^{+1}_{\mathrm{out}}}^{2}+\braket{\hat{p}^{+1}_{\mathrm{out}}}^{2}}|$ is proportional to $|J_1(\rho)|$. Since the modulation-depth $\delta \theta$ and hence the modulation parameter $\rho$ upscales as the number of optimally cascaded antennas are increased, it results in a larger  $|J_1(\rho)|$, and hence larger $d$.

\begin{figure}[ht]
    \centering
    \includegraphics[width=\linewidth]{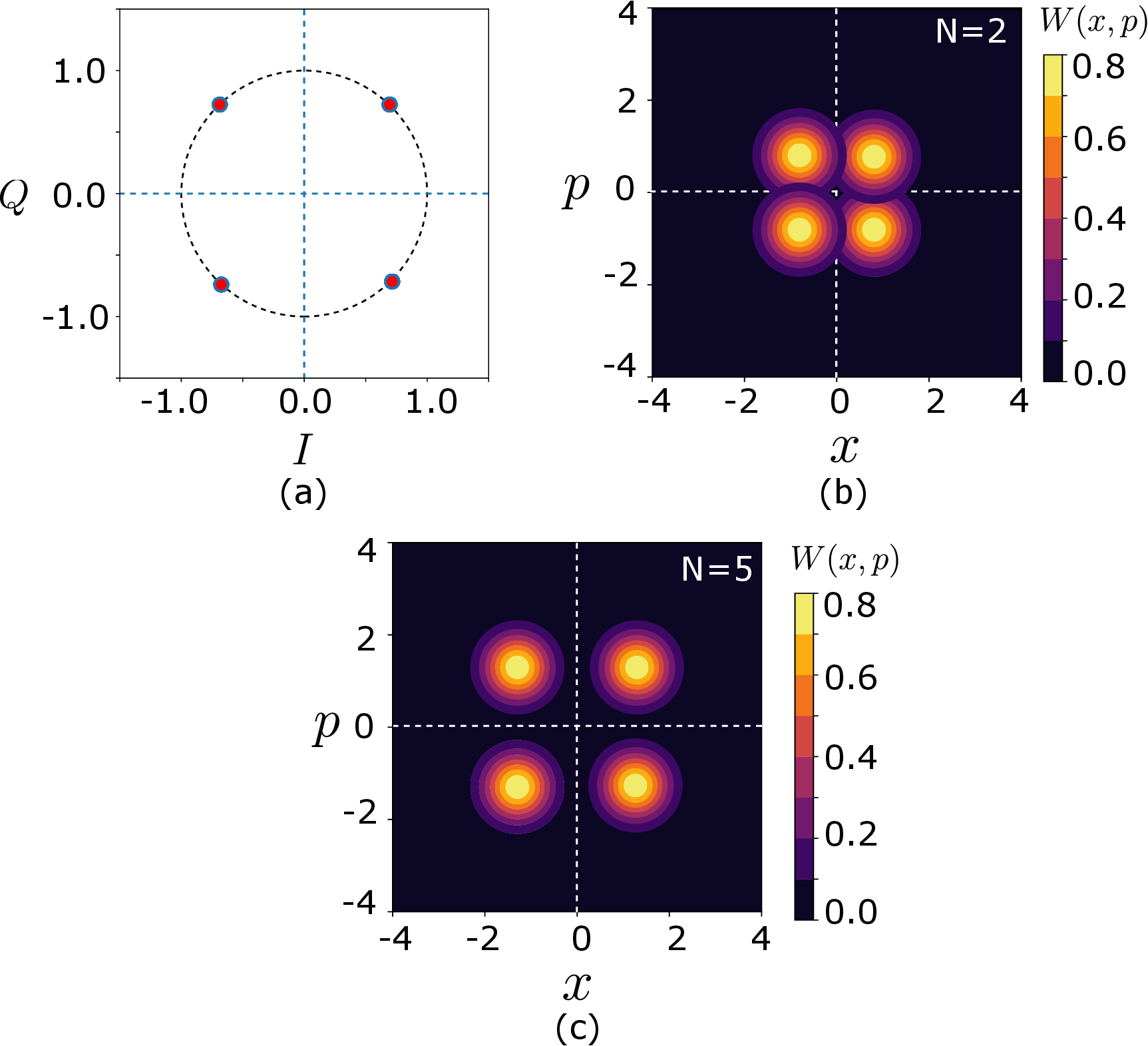}
    \caption{(a) Constellation diagram of a QPSK signal encoded in a 30GHz classical mmWave signal of E-field strength $|E_{\mathrm{m}}|$=100V/m. The corresponding Wigner distribution of the quantum optical coherent state composed of a mean photon number pf $\tilde{n}_{ph}$=4 modulated in a seamless converter composed of (b) N=2, and (c) N=5 number of optimally cascaded antennas.}
    \label{Figure3}
\end{figure}

\begin{figure}[ht]
    \centering
    \includegraphics[width=\linewidth]{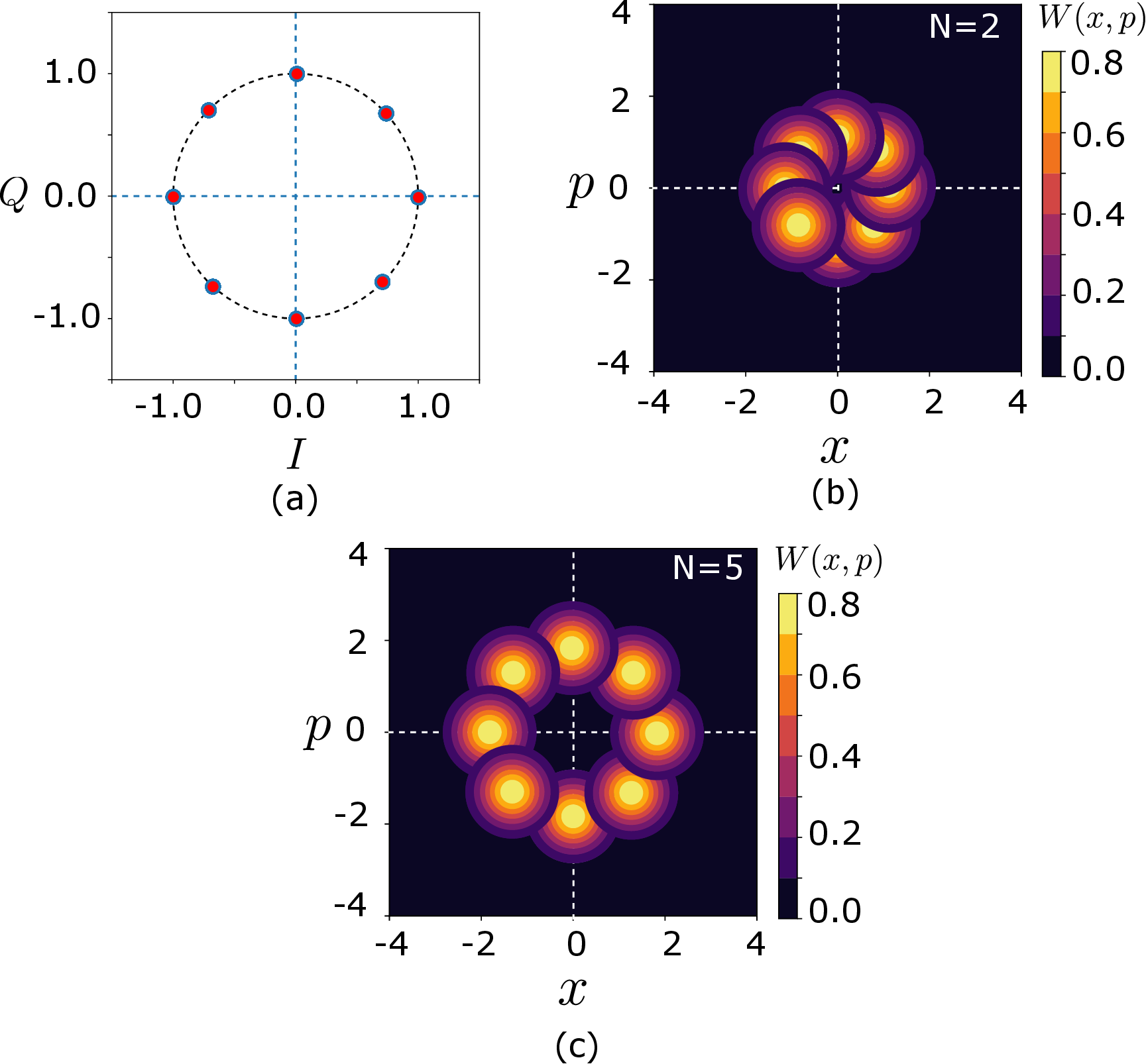}
    \caption{(a) Constellation diagram of a 8-PSK signal encoded in a 30GHz classical mmWave signal of E-field strength $|E_{\mathrm{m}}|$=100V/m. The corresponding Wigner distribution of the quantum optical coherent state composed of a mean photon number of $\tilde{n}_{ph}$=4 modulated in a seamless converter composed of (b) N=2, and (c) N=5 number of optimally cascaded antennas.}
    \label{Figure4}
\end{figure}


We now extend our discussion to investigate optical Wigner-space conditioning through classical mmWave digital phase-shift-keying in Figure~\ref{Figure3}. The constellation diagram of a QPSK signal embedded in a 30GHz mmWave carrier of E-field strength $|E_{\mathrm{m}}|$=100V/m is shown in Fig.~\ref{Figure3}(a). The four data points in the QPSK constellation are associated with the phase-levels $b_i$=$\{45^{\circ}, 135^{\circ}, 225^{\circ}, 315^{\circ}\}$. Each phase-level baseband $b_i$ from mmWave constellation gets translated to distinct mean phase-level $\phi_i$ in the Wigner-space of the modulated optical coherent-state, as shown in Fig.~\ref{Figure3}(b). It must be reiterated that for the computation we have considered optimal design, due to which the modulation-depth is linearly proportional to the number of antennas in the array, and constellation-rotation angle $r$ is 
 $0^{\circ}$. It must however be noted from Fig.~\ref{Figure3}(b) that the Wigner distributions of the modulated optical coherent-states corresponding to different symbols in mmWave constellation tend to overlap.  This is primarily due to the fact that for N=2 number of antennas in the array, the modulation-depth is not large enough. Consequently, the first-order sideband amplitude of the modulated coherent state, and hence its mean radial distance from the origin of the phase-space is not large enough  to completely avoid intersymbol overlap. On the other hand, when the number of antennas in the array is increased to N=5, the modulation-depth, and consequently the first-order sideband amplitude of the modulated coherent state is relatively large enough to completely avoid intersymbol overlap, as shown in Fig.~\ref{Figure3}(c).

A similar analysis for classical mmWave 8-PSK reception is shown in Fig.~\ref{Figure4}. The constellation diagram of a 8-PSK signal with eight distinct phase-levels $b_i$=$\{0^{\circ}, 45^{\circ}, 90^{\circ}, 135^{\circ}, 180^{\circ}, 225^{\circ}, 270^{\circ}, 315^{\circ}\}$ are shown in Fig.~\ref{Figure4}(a). The correspondingly conditioned Wigner distributions of the modulated optical coherent states in a seamless converter of N=2 and N=5 optimally cascaded antennas are shown in Fig.~\ref{Figure4}(b) and Fig.~\ref{Figure4}(c), respectively. Even in this case, it can be observed that the extent of intersymbol overlap in the quantum optical Wigner distributions decreases as N is increased from 2 to 5. However, compared to the QPSK case, the intersymbol overlap in the 8-PSK case is higher due to the presence of a relatively larger number of data points in the received classical constellation.



    \label{Figure6}

For the computation of all the results, we have considered the baseband symbol phase $b_i$ to be encapsulated in a 30GHz classical mmWave signal of E-field strength $|E_\mathrm{m}|$=100V/m. We further consider the 1555nm optical coherent-state to be composed of mean photon numbers $\Tilde{n}_{ph}$=4, having mean E-field magnitude $|\alpha|$=$\sqrt{\Tilde{n}_{ph}}$=2. We have also assumed the mean initial phase of the coherent-state to be $\phi$=$0^{\circ}$. 

In the upcoming section, we will extend the investigation to the case of optical squeezed states.

\section{Conditioning optical squeezed states}

An  optical displaced squeezed state $\ket{\alpha_{\mathrm{sq}}}$ can be defined as a displaced squeezed vacuum state, and mathematically described as follows \cite{scully1997quantum},
\begin{equation}
  \ket{\alpha_{\mathrm{sq}}} = \hat{D}(\alpha) \hat{S}(r) \ket{0}
\label{Eq23}
\end{equation}
where $\hat{D}(\alpha)$=$\exp{(\hat{a}^{\dagger} \alpha - \hat{a} \alpha^{*})}$ is the displacement operator, $\hat{S}(r)$=$\exp{(0.5r(\hat{a}^{\dagger 2} - \hat{a}^{2})})$ is the squeezing operator, $r$ is the squeezing parameter, and $\ket{0}$ represents a vacuum state \cite{scully1997quantum}. Also here, like that of a coherent state, $\alpha$ corresponds to the mean complex amplitude of the E-field associated with the squeezed state $\ket{\alpha_{\mathrm{sq}}}$. 

\begin{figure*}[ht]
    \centering
    \includegraphics[width=\linewidth]{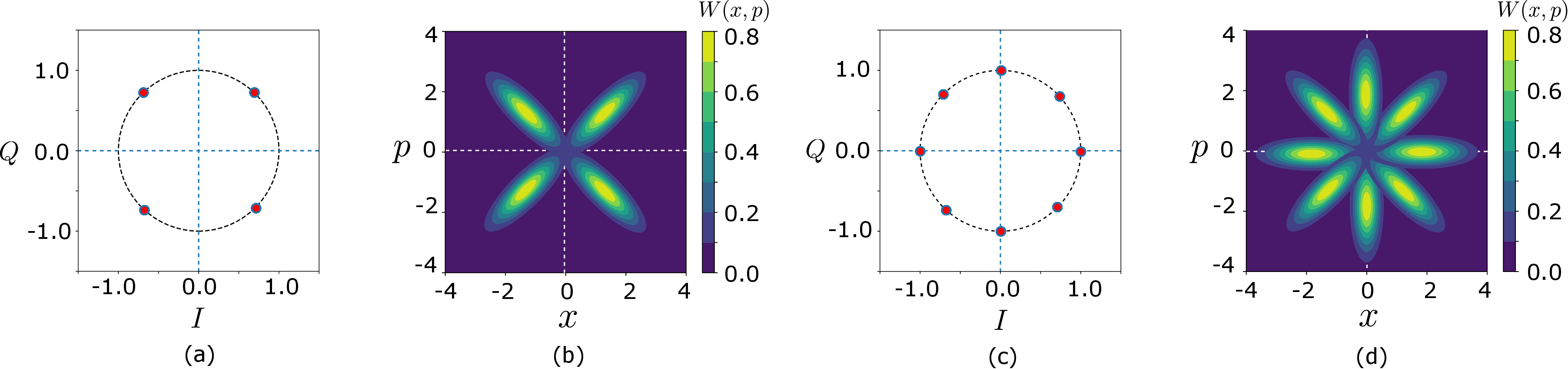}
    \caption{(a) Constellation diagram of a QPSK signal encoded in a 30GHz classical mmWave signal of E-field strength $|E_{\mathrm{m}}|$=100V/m. (b)  The Wigner distribution of the correspondingly modulated optical squeezed-state. (c) Constellation diagram of a 8-PSK signal encoded in a 30GHz classical mmWave signal of E-field strength $|E_{\mathrm{m}}|$=100V/m. (d)  The Wigner distribution of the correspondingly modulated optical squeezed-state. Here, the optical squeezed state is considered to be composed of mean photon numbers $\tilde{n}_{ph}$=4, with squeezing parameter $r$=0.6, and the number of optimally cascaded antennas in the seamless converter is $N$=5.}
    \label{Figure7}
\end{figure*}

We now consider the scenario in which the displaced optical squeezed state $\ket{\alpha_{\mathrm{sq}}}$ is modulated by a classical mmWave signal received by the seamless converter. We consider the unmodulated quantum optical state at the converter input to be squeezed in $p$ by a factor of $e^{-r}$ and anti-squeezed in $x$ by a factor of $e^{r}$. The Wigner function $W(x,p)$ associated with the first-order sideband of the modulated displaced squeezed state can be expressed as follows,
\begin{widetext}
\begin{equation}
W(x,p) = \frac{2}{\pi} \exp{\big(-2(e^{-2r}(x-\braket{\hat{x}^{+1}_{\mathrm{out}}}_{})^{2}+ \nonumber \\
e^{2r}(p-\braket{\hat{p}^{+1}_{\mathrm{out}}}_{})^{2}})\big)
\label{EqWsq}
\end{equation}
\end{widetext}
where $\braket{\hat{x}^{+1}_{\mathrm{out}}}_{}$ and $\braket{\hat{p}^{+1}_{\mathrm{out}}}_{}$ is the mean $\hat{x}$ and $\hat{p}$ associated with the first-order sideband of the modulated displaced squeezed state $\ket{\alpha_{\mathrm{sq}}}$, respectively. 
\begin{figure*}[ht]
    \centering
    \includegraphics[width=\linewidth]{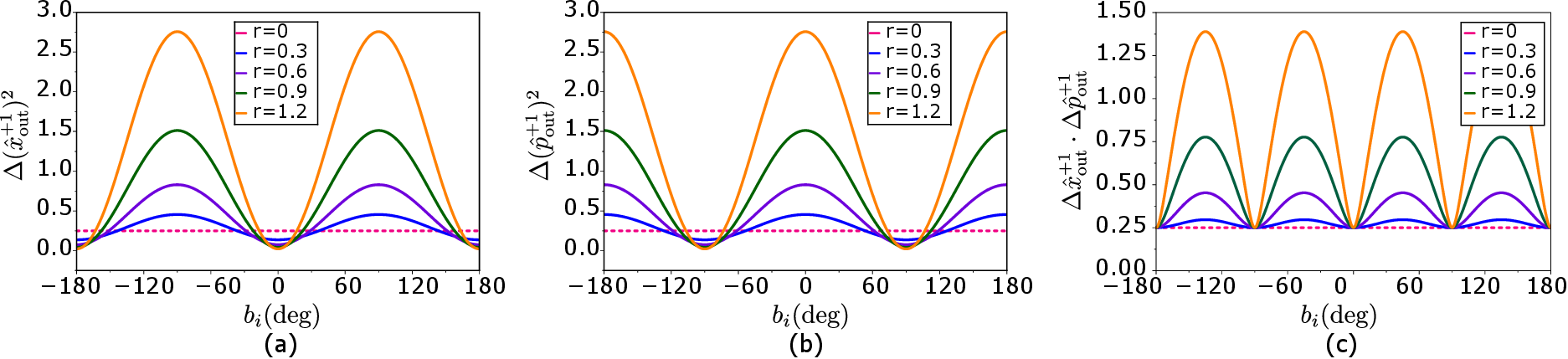}
    \caption{Variation in quadrature fluctuations (a) $\Delta ({\hat{x}^{+1}_{\mathrm{out}}})^{2}$,  (b) $\Delta ({\hat{p}^{+1}_{\mathrm{out}}})^{2}$, and (c) Quadrature uncertainty product $\Delta{\hat{x}^{+1}_{\mathrm{out}}} \cdot \Delta{\hat{p}^{+1}_{\mathrm{out}}}$ of the modulated displaced optical squeezed state ($\tilde{n}_{ph}$=4) corresponding to different squeezing levels $r$, observed by sweeping the baseband symbol phase $b_i$ encoded in the 30GHz classical mmWave signal of E-field strength $|E_\mathrm{m}|$=100V/m received by the seamless converter of N=5 number of antennas.}
    \label{Figure10}
\end{figure*}

Following a similar procedure as the previous subsection, the Wigner function $W(x,p)$ of the modulated squeezed-state can be derived to be,
\begin{widetext}
\begin{equation}
\begin{split}
 W(x,p) =  \frac{2}{\pi} \exp{\big(-2(e^{-2r}(x-\Re{\{\alpha\}} J_1(\rho)\cos{\phi_i}  - \Im{\{\alpha\}} J_1(\rho)\sin{\phi_i})^{2}+e^{2r}(p+\Re{\{\alpha\}} J_1(\rho)\sin{\phi_i}- \Im{\{\alpha\}} J_1(\rho)\cos{\phi_i})^{2})\big)}
\end{split}
\label{Eq25sq}
\end{equation}
\end{widetext}

Figure~\ref{Figure7} shows Wigner-space conditioning of displaced optical squeezed states through classical mmWave digital phase-shift-keying. The constellation diagram of a mmWave QPSK constellation is shown in Fig.~\ref{Figure7}(a). The Wigner distribution of the correspondingly modulated optical squeezed state is shown in Fig.~\ref{Figure7}(b). It can be observed that each baseband phase-level $b_i$ embedded in the received mmWave QPSK constellation gets mapped to distinct mean phase-levels $\phi_i$ in the Wigner-space of the modulated optical squeezed state. Very much like the previously discussed case of coherent state conditioning, we can observe linear mapping of phase from the classical mmWave constellation to the Wigner-space of the optical squeezed state. For the computation, we have considered the initial mean phase of the squeezed state to be $0^{\circ}$, squeezing parameter to be $r$=0.6, and the number of optimally cascaded antennas in the converter to be $N$=5. Figure~\ref{Figure7}(c)-(d) show classical mmWave constellation to quantum optical Wigner distribution mapping when the optical squeezed-state is modulated by a 8-PSK mmWave signal received by the seamless converter. It can be observed that each of the eight baseband phase-levels embedded in the received 8-PSK mmWave constellation get mapped to eight distinct mean phase-levels in the Wigner-space of the modulated optical squeezed state.  In this case, a relatively smaller inter-symbol overlap can be observed compared to the previously discussed case of coherent-state encoding with 8-PSK in Fig.~\ref{Figure4}(c).

It is important to highlight that sweeping the baseband symbol phase $b_i$ encoded in the modulating mmWave signal modifies the quadrature variances of the modulating squeezed-states, in addition to modifying the mean quadratures. This is because the Wigner distribution of squeezed states  have elliptical Gaussian symmetry around the mean. For an unmodulated quantum optical state initially anti-squeezed in $x$ and squeezed in $p$, the variances in the respective quadratures are $\Delta \hat{x}^{2}_{\ket{\alpha_{\mathrm{sqo}}}}$=$1/4 e^{2r}$ and $\Delta \hat{p}^{2}_{\ket{\alpha_{\mathrm{sqo}}}}$=$1/4 e^{-2r}$. Since mmWave phase-shifting rotates the mean location of the squeezed state in the Wigner-space, the projection of its quasi-probability distribution on the quadrature axes modified as a consequence. This can be verified by analytically deriving the quadrature variance $\Delta ({\hat{x}^{+1}_{\mathrm{out}}})^{2}$ of the modulated displaced optical squeezed state as follows,
\begin{equation}
\begin{split}
\Delta ({\hat{x}^{+1}_{\mathrm{out}}})^{2} & = \bra{\alpha_{\mathrm{sq}}} (\hat{x}^{+1}_{\mathrm{out}} )^{2}{\ket{\alpha_{\mathrm{sq}}}} -\bra{\alpha_{\mathrm{sq}}} \hat{x}^{+1}_{\mathrm{out}} {\ket{\alpha_{\mathrm{sq}}}}^{2}\\
& =  \Delta \hat{x}^{2}_{\ket{\alpha_{\mathrm{sqo}}}} \cos^{2}{\phi_i} + \Delta \hat{p}^{2}_{\ket{\alpha_{\mathrm{sqo}}}} \sin^{2}{\phi_i}
\end{split}
\label{Eqdelx}
\end{equation}
Similarly, the quadrature variance $\Delta ({\hat{p}^{+1}_{\mathrm{out}}})^{2}$ can be derived to be,
\begin{equation}
\begin{split}
\Delta ({\hat{p}^{+1}_{\mathrm{out}}})^{2} & = \bra{\alpha_{\mathrm{sq}}} (\hat{p}^{+1}_{\mathrm{out}} )^{2}{\ket{\alpha_{\mathrm{sq}}}} -\bra{\alpha_{\mathrm{sq}}} \hat{p}^{+1}_{\mathrm{out}} {\ket{\alpha_{\mathrm{sq}}}}^{2}\\
& =  \Delta \hat{x}^{2}_{\ket{\alpha_{\mathrm{sqo}}}} \sin^{2}{\phi_i} + \Delta \hat{p}^{2}_{\ket{\alpha_{\mathrm{sqo}}}} \cos^{2}{\phi_i}
\end{split}
\label{Eqdelp}
\end{equation}
From Eq.~(\ref{Eqdelx}) and Eq.~(\ref{Eqdelp}), it can be a discerned that the quadrature variances $\Delta ({\hat{x}^{+1}_{\mathrm{out}}})^{2}$ and $\Delta ({\hat{p}^{+1}_{\mathrm{out}}})^{2}$ are functions of 
the mean modulated quantum optical phase $\phi_i$. Since we already know that the  baseband symbol phase $b_i$ governs the mean modulated quantum optical phase $\phi_i$, it can be concluded that $\Delta ({\hat{x}^{+1}_{\mathrm{out}}})^{2}$ and $\Delta ({\hat{p}^{+1}_{\mathrm{out}}})^{2}$ are functions of $b_i$.

Figure~\ref{Figure10}(a) and (b) show the variation in $\Delta ({\hat{x}^{+1}_{\mathrm{out}}})^{2}$ and $\Delta ({\hat{p}^{+1}_{\mathrm{out}}})^{2}$, respectively, of the modulated displaced optical squeezed state, observed by varying the baseband symbol phase from $b_i$=$-180^{\circ}$ to $b_i$=$180^{\circ}$, corresponding to different squeezing levels $r$. It can be discerned from Fig.~\ref{Figure10}(a)-(b) that $\Delta ({\hat{x}^{+1}_{\mathrm{out}}})^{2}$ and $\Delta ({\hat{p}^{+1}_{\mathrm{out}}})^{2}$ are sinusoidal functions of the baseband symbol phase $b_i$,which can be analytically verified from Eq.~(\ref{Eqdelx}) and Eq.~(\ref{Eqdelp}). It can be noticed that when the squeezing-level $r$=0, we are essentially referring to a coherent-state that has equal quadrature variances i.e. $\hat{x}^{2}_{\ket{\alpha_{\mathrm{o}}}}$=$\hat{p}^{2}_{\ket{\alpha_{\mathrm{o}}}}$=0.25. Since the Wigner-distribution of coherent states have circularly symmetric Gaussian spread around mean location in phase-space, the rotation induced by baseband phase-shift-keying does not impact it. This is why $\Delta ({\hat{x}^{+1}_{\mathrm{out}}})^{2}$ and $\Delta ({\hat{p}^{+1}_{\mathrm{out}}})^{2}$ remain constant, even when when $b_i$ is varied, as shown in Fig.~\ref{Figure10}(a) and (b). Naturally, it can also be shown that the quadrature uncertainty product $\Delta{\hat{x}^{+1}_{\mathrm{out}}} \cdot \Delta{\hat{p}^{+1}_{\mathrm{out}}}$ of the modulated squeezed state can be manipulated by sweeping the baseband data phase $b_i$. This can be analytically supported by the equation mentioned below,
\begin{widetext}
   \begin{equation}
    \Delta{\hat{x}^{+1}_{\mathrm{out}}} \cdot \Delta{\hat{p}^{+1}_{\mathrm{out}}} = \sqrt{(\Delta \hat{x}^{2}_{\ket{\alpha_{\mathrm{sqo}}}}-\Delta \hat{p}^{2}_{\ket{\alpha_{\mathrm{sqo}}}})\ \sin^{2}{\phi_i}\ \cos^{2}{\phi_i} + \Delta \hat{x}^{2}_{\ket{\alpha_{\mathrm{sqo}}}} \cdot \Delta \hat{p}^{2}_{\ket{\alpha_{\mathrm{sqo}}}} }
\label{Eq_Uncertainty}
\end{equation}
\end{widetext}


Figure~\ref{Figure10}(c) shows variation in the quadrature uncertainty $\Delta{\hat{x}^{+1}_{\mathrm{out}}} \cdot \Delta{\hat{p}^{+1}_{\mathrm{out}}}$ of the modulated squeezed state obtained by sweeping the classical mmWave phase $b_i \in [-180^{\circ}, 180^{\circ}]$. It can be observed that $\Delta{\hat{x}^{+1}_{\mathrm{out}}} \cdot \Delta{\hat{p}^{+1}_{\mathrm{out}}}$ does not go below the lower limit $\hat{x}_{\ket{\alpha_{\mathrm{sqo}}}} \cdot \hat{p}_{\ket{\alpha_{\mathrm{sqo}}}}$=1/4 set by the Heisenberg's uncertainty principle. This shows that the mmWave-embedded baseband symbol phase not only determines the mean displacement of the modulated squeezed state, but also enables symbol-dependent quadrature fluctuation engineering. This enables simultaneous imprinting of information across multiple quantum photonic degrees of freedom, and can have a plethora of applications in quantum signal processing, communication, and hybrid classical-quantum networking in the near future.

\section{Conclusion and discussion}

This work presents a Heisenberg-picture model for contactless conditioning of CV quantum states of light through digitally modulated classical mmWave beams. A direct mapping between classical mmWave phase variation and the resulting Wigner-space evolution of modulated optical coherent and squeezed states is established. Interestingly, it is revealed that quadrature uncertainty of squeezed states can be modulated through classical mmWave phase-shift-keying, enabling contactless imprinting of information onto multiple quantum-optical degrees of freedom. 

From a technological perspective, the presented theoretical framework provides a foundation for replacing conventional on-chip RF interconnects with directive wireless mmWave or THz interfaces. This can potentially reduce interconnect loss, crosstalk, hardware complexity, and power consumption in integrated quantum photonic platforms. Practical realization will however require further improvements in wireless-to-photonic coupling efficiency, device integration, thermal stability, fabrication tolerances, and availability of efficient quantum photonic sources and detector. Nevertheless, experimental implementation can leverage established state-of-art frequency-division multiplexing and spatial-division multiplexing techniques to enable parallel addressing of multiple photonic devices or quantum channels. Such integration of modern wireless communication technology with emerging quantum photonics provides a feasible pathway toward scalable, contactless quantum photonic information processing and networking.




\providecommand{\noopsort}[1]{}\providecommand{\singleletter}[1]{#1}%

\end{document}